\documentclass[11pt]{article}
 \pdfoutput=1
\usepackage{graphicx,wrapfig,color}
\usepackage{subfig}
\usepackage{setspace}
\usepackage{epsfig}
\usepackage{graphicx}
\usepackage{multicol}
\usepackage{amsfonts}
\usepackage{latexsym}
\usepackage{amsmath,amssymb}
\usepackage{enumerate}
\usepackage{hyperref}
\hypersetup{
    colorlinks=flase,
    linkcolor=black,
    filecolor=magenta,      
    urlcolor=blue,
}
\usepackage{cite}
\usepackage{lipsum}
\usepackage{mwe}
\usepackage{subfig}
\newcommand{\ben}{\begin{enumerate}}
\newcommand{\een}{\end{enumerate}}

\newcommand{\bea}{\begin{eqnarray}}
\newcommand{\eea}{\end{eqnarray}}
\newcommand{\be}{\begin{equation}}
\def\bel#1{\begin{equation} \label{#1}}
\newcommand{\ee}{\end{equation}}
\newcommand{\bi}{\begin{itemize}}
\newcommand{\ei}{\end{itemize}}
\newcommand{\ba}{\begin{align}}
\newcommand{\ea}{\end{align}}
\def\bel#1{\begin{equation} \label{#1}}

\def\be{\begin{equation}}
\def\ee{\end{equation}}
\def\bea{\begin{eqnarray}}
\def\eea{\end{eqnarray}}

\def\ltap{\ \raise.3ex\hbox{$<$\kern-.75em\lower1ex\hbox{$\sim$}}\ }
\def\gtap{\ \raise.3ex\hbox{$>$\kern-.75em\lower1ex\hbox{$\sim$}}\ }
\def\gl{\ \raise.5ex\hbox{$>$}\kern-.8em\lower.5ex\hbox{$<$}\ }
\def\roughly#1{\raise.3ex\hbox{$#1$\kern-.75em\lower1ex\hbox{$\sim$}}}

\newcommand{\comments}[1]{}

\usepackage{color}
\definecolor{cblue}{RGB}{100,5,255}
\definecolor{cred}{RGB}{255,50,40} 
\definecolor{cgreen}{RGB}{40,255,40} 
\definecolor{corange}{RGB}{250,200,40}

\begin{document}

\begin{titlepage}
\vskip 1 cm
\begin{center}
{\Large \bf Constraining Non-thermal Dark Matter by CMB} 
%
\vskip 1.5cm  
{ 
Rouzbeh Allahverdi$^{\dagger}$, Koushik Dutta$^{\ddagger}$ and Anshuman Maharana$^{*}$  
\let\thefootnote\relax\footnotetext{ \hspace{-0.5cm} E-mail: {$\mathtt{rouzbeh@unm.edu, koushik.dutta@saha.ac.in, anshumanmaharana@hri.res.in } $}}
}
\vskip 0.9 cm

{\textsl{$^{\dagger}$ Department of Physics and Astronomy \\
University of New Mexico \\
1919 Lomas Blvd. NE \\
Albuquerque NM 87131-0001, USA}
}
\vskip 0.6 cm
{\textsl{$^{\ddagger}$Theory Division, \\
 Saha Institute of Nuclear Physics, \\
 HBNI, 1/AF Salt Lake,  \\
 Kolkata - 700064, India.}
}
\vskip 0.6cm
{\textsl{
$^{*}$Harish Chandra Research Intitute, \\
HBNI, Chattnag Road, Jhunsi,\\
Allahabad -  211019, India.\\}
}

\vskip 0.6 cm

%
%
\end{center}

\vskip 0.6cm

\begin{abstract}
A period of early matter domination can give rise to the correct dark matter abundance for a broad range of dark matter annihilation rate $\langle \sigma_{\rm ann} v \rangle_{\rm f}$. Here, we examine this scenario for situations where $\langle \sigma_{\rm ann} v \rangle_{\rm f}$ is below the nominal value for thermal dark matter $3 \times 10^{-26}$ cm$^3$ s$^{-1}$ as possibly indicated by some recent experiments. We show that obtaining the correct relic abundance sets a lower bound on the duration of early matter domination era in this case. On the other hand, provided that the post-inflationary universe has an equation of state characterized by $w \leq 1/3$, the requirement that the scalar spectral index $n_s$ be within the observationally allowed range limits the duration of this epoch from above. By combining these considerations, we show that the current and future cosmic microwave background experiments can tightly constrain the parameter space for this scenario. In particular, models of inflation with a tensor-to-scalar ratio below ${\cal O}(0.01)$ may disfavor non-thermal supersymmetric dark matter from a modulus-driven early matter domination epoch.    

\end{abstract}

\vspace{3.0cm}

\end{titlepage}
\pagestyle{plain}
\setcounter{page}{1}
\newcounter{bean}
\baselineskip18pt
%



\section{Introduction}

Despite various lines of evidence for the existence of dark matter (DM) in the universe~\cite{BHS}, its identity remains as a major problem at the interface of cosmology and particle physics. Weakly interacting massive particles (WIMPs) are promising candidates for DM and are the main focus of direct, indirect, and collider searches that are currently underway to discover DM. A nice mechanism for obtaining the correct abundance for WIMP DM is the "WIMP miracle", which assumes that the universe was in a radiation-dominated (RD) phase at temperatures about the DM mass $m_\chi$. The DM relic abundance in this picture is set when the annihilation rate of DM particles drops below the Hubble expansion rate, called "thermal freeze-out", and matches the observed value if the annihilation rate takes the nominal value $\langle \sigma_{\rm ann} v \rangle_{\rm f} = 3 \times 10^{-26}$ cm$^3$ s$^{-1}$. However, the WIMP miracle has come under increasing scrutiny by the recent experimental data. For example, Fermi-LAT's results from observations of dwarf spheroidal galaxies~\cite{fermi1} and newly discovered Milky Way satellites~\cite{fermi2} place an upper bound on the DM annihilation rate that is below the nominal value required for the WIMP miracle for a range of DM masses. A recent analysis~\cite{Beacom} shows that in models where DM annihilation is dominated by $S$-wave processes, thermal DM with a mass below 20 GeV is ruled out in a model-independent way, while for certain annihilation channels this will be the case for masses up to 100 GeV. This implies that thermal freeze-out in a RD universe would lead to overproduction of DM within the corresponding mass range (unless there is $P$-wave annihilation or co-annihilation, in which cases the WIMP miracle condition and the indirect detection limits may be both satisfied).

However, the situation can change in a non-standard thermal history where the universe is not RD at the time of freeze-out~\cite{KT}. In particular, it is known that an epoch of early matter domination (EMD) that ends before the onset of big bang nucleosynthesis (BBN) can accommodate DM annihilation rate below $3 \times 10^{-26}$ cm$^3$ s$^{-1}$~\cite{GKR, KMY,GG, ADS}. Interestingly, an EMD era is a generic feature of an important class of early universe models arising from string theory constructions (for a review, see~\cite{KSW}). In these models, the modulus fields are displaced from the minimum of their potential during inflation due to misalignment~\cite{cmp}, and dominate the energy density of the post-inflationary universe due to their long lifetime. Late decay of moduli reheats the universe to temperatures below the freeze-out temperature $T_{\rm f}$, thereby rendering the WIMP miracle irrelevant in this framework.

The presence of an EMD epoch in the early universe typically decreases the number of e-foldings between the horizon exit of observationally relevant cosmological perturbations and the end of inflation~\cite{LL}. However, such a change also affects inflationary predictions for the scalar spectral index $n_s$ and the tensor-to-scalar ratio $r$~\cite{EGOW,DDM}. Furthermore, \cite{EGOW} discuses some connections between supersymmetry, non-thermal DM and precision cosmology.\footnote{For some related work along this line, see \cite{related,infreh}.} This implies that cosmic microwave background (CMB) experiments may be used to constrain the non-thermal DM production from an epoch of EMD. 

In this paper, we explore this issue for the case with small DM annihilation rate $\langle \sigma_{\rm ann} v \rangle_{\rm f} < 3 \times 10^{-26}$ cm$^3$ s$^{-1}$. We consider various contributions to the DM abundance and obtain an absolute lower bound on the duration of the EMD era without making explicit reference to its particle physics origin. We contrast this with the upper bounds derived from the $n_s$ considerations for representative models of inflation that are compatible with the latest Planck results~\cite{planckinflation}. We show that the current~\cite{planckinflation} and future~\cite{next1} CMB experiments can tightly constrain the parameter space for an epoch of EMD. In particular, provided that the post-inflationary universe has an equation of state characterized by $w \leq 1/3$, a typical modulus-driven EMD period as the origin of non-thermal supersymmetric DM may be disfavored for inflationary models with $r \lesssim {\cal O}(0.01)$.

The rest of this paper is organized as follows. In Section II, we discuss various contributions to DM production during EMD and derive an absolute lower bound on the duration of this period. In Section III, we discuss connection to inflationary observables using a parametrization of $n_s$ and $r$ that holds for a large number of inflationary models compatible with the latest observational data. In Section IV, we present our main results along with some disucusion. Finally, we conclude the paper in Section V.

\section{Non-thermal Dark Matter from Early Matter Domination}

WIMP miracle, while being a simple and predictive scenario, is coming under increasing pressure {by experiments (for example, see~\cite{fermi1,fermi2}). This motivates studying alternative scenarios of DM production, which has recently witnessed an increasing activity (for a review, see~\cite{Howie}). A particularly attractive scenario, as mentioned above, is non-thermal DM production during an epoch of EMD. 

An era of EMD can arise from oscillations of a long-lived scalar field. Consider a scalar field $\phi$ with mass $m_\phi$ and decay width $\Gamma_\phi$. Such a field is typically displaced from the true minimum of its potential during inflation. It starts oscillating about the minimum with an initial amplitude $\phi_0$ when the Hubble expansion rate is $H_{\rm osc} \simeq m_\phi$. The fractional energy density of $\phi$ at the onset of oscillations is given by $\alpha_0 \simeq (\phi_0/M_{\rm P})^2$. Since these oscillations behave like matter, their fractional energy density increases as $\alpha(t) \propto a(t) \propto H^{-1/2}$ in a RD universe. The oscillations start to dominate the energy density of the universe when $\alpha(t) \simeq 1$. This happens when the Hubble expansion rate is $H_{\rm dom} \simeq \alpha^{2}_0 m_\phi $,\footnote{Henceforth, we use $\alpha_0$ instead of $(\phi_0/M_{\rm P})^2$. This will  keep our discussion more general as it can be readily applied to situations where the EMD period is driven by non-relativistic quanta produced in the post-inflationary universe instead of coherent oscillations of a scalar field (for example, see~\cite{Hooper}).} at which time the universe enters an epoch of EMD. Oscillations eventually decay when the Hubble expansion rate is $H_{\rm R} \simeq \Gamma_\phi$, resulting in a RD universe with the following reheat temperature:
\be \label{treh}
T_{\rm R} \simeq \left({90 \over \pi^2 g_{*,{\rm R}}}\right)^{1/4} \sqrt{\Gamma_\phi M_{\rm P}} , 
\ee
where$g_{*,{\rm R}}$ is the number of relativisitic degrees of freedom at at temperature $T_{\rm R}$.~\footnote{Here we assume that $\phi$ decays perturbatively, which is justified if its couplings to other fields are sufficiently small and its potential is not very steep.} 

Decay of $\phi$ is a continuous process and, assuming that decay products are relativistic and thermalize immediately, it results in a thermal bath during EMD whose instantaneous temperature $T$ follows (for $H_{\rm R} \ll H \ll H_{\rm dom}$)~\cite{GKR}:
\be \label{tinst}
T = \left({6 \sqrt{g_{*,{\rm R}}} \over 5 g_*}\right)^{1/4} \left({30 \over \pi^2}\right)^{1/8} \left(H T^2_{\rm R} M_{\rm P}\right)^{1/4} , 
\ee
where $g_*$ denotes the number of relativistic degrees of freedom at temperature $T$. This thermal bath is subdominant when $H \gg H_{\rm R}$, but carries the entire energy density upon completion of $\phi$ decay at $H_{\rm R}$. We see from Eq.~(\ref{treh}) and Eq.~(\ref{tinst}) that $T \gg T_{\rm R}$ for $H \gg \Gamma_\phi$, which implies DM production from thermal porcesses is possible during EMD. For small DM annihilation rates, $\langle \sigma_{\rm ann} v \rangle_{\rm f} < 3 \times 10^{-26}$ cm$^3$ s$^{-1}$, the correct DM abundance can be obtained via thermal freeze-out/freeze-in during EMD~\cite{GKR}, or via direct production at the very end of the EMD epoch~\cite{KMY,GG, ADS}. 
\footnote{For large annihilation rates, $\langle \sigma_{\rm ann} v \rangle_{\rm f} > 3 \times 10^{-26}$ cm$^3$ s$^{-1}$, direct production at the end of EMD (with or without residual annihilation) is the only way to yield the correct DM abundance~\cite{MR}. The implications of non-thermal DM for CMB observables in this case are studied in~\cite{EGOW}, which discusses production of supersymmetric WIMPs from an epoch of EMD driven by a modulus field.} 

We now consider contributions from all of the above processes and derive lower limits on the duration of the EMD era by requiring that it gives rise to the correct DM abundance. We adopt a model-independent approach to the EMD phase, which does not make referecne to its explicit particle physics origin, based  on two parameters $m_\phi$ and $T_{\rm R}$.   
\vskip 2mm
\noindent
{$\bullet$ {\bf Decays at the end of EMD:}} 
Decay of $\phi$ oscillations reheat the universe and produce DM particles with the following abundance~\cite{GG, ADS}:
\be \label{dec}
\left({n_\chi \over s}\right)_{\rm dec} = {3 T_{\rm R} \over 4 m_\phi} ~ {\rm Br}_{\phi \rightarrow \chi} .
\ee
Here ${\rm Br}_{\phi \rightarrow \chi}$ denotes the branching fraction for production of DM particles per $\phi$ decay. It includes both direct and indirect production of DM from $\phi$ decay. For example, in supersymmetric models, ${\rm Br}_{\phi \rightarrow \chi}$ is the branching fraction for decay to $R$-parity odd particles, with the heavier ones ending up in $\chi$ via cascade decays.  

In order to match the observed DM abundance, we need to have:
\be \label{decdens}
{3 T_{\rm R} \over 4 m_\phi} ~ {\rm Br}_{\phi \rightarrow \chi} \simeq 5 \times 10^{-10} ~ \left({ 1 ~ {\rm GeV} \over m_\chi}\right).
\ee
After using Eq.~(\ref{treh}), the relation $m_\phi \simeq \alpha^{-2}_0 H_{\rm dom}$, and the fact that $T_{\rm R}$ must be smaller than the DM freeze-out/freeze-in tempeature during EMD $T_{\rm f} \lesssim m_\chi/5$ (as discussed below), we arrive at:
\be \label{decbound}
{H_{\rm dom} \over H_{\rm R}} \gtrsim 10^{10} \left({90 \over \pi^2 g_{*,{\rm R}}}\right)^{1/2} \left({M_{\rm P} \over 1 ~ {\rm GeV}}\right) ~ \alpha^2_0 ~ {\rm Br}_{\phi \rightarrow \chi}.
\ee
%

%
\vskip 2mm
\noindent
{$\bullet$ {\bf Freeze-out during EMD:}} 
DM freeze-out temperature in the EMD phase typically lies in the range $m_\chi/25 \lesssim T_{\rm f} \lesssim m_\chi/5$, with the exact value depending on the annihilation rate. Since $\langle \sigma_{\rm ann} v \rangle _{\rm f} < 3 \times 10^{-26}$ cm$^3$ s$^{-1}$, DM is overproduced initially but its abundance is diluted as $\phi$ decay keeps injecting entropy until the transition to RD completes. The DM relic abundance due to freeze-out is given by~\cite{GKR}:  
\be \label{fodens}
\Omega_{\chi} h^2 \simeq 1.6 \times 10^{-4} {\sqrt{g_{*,{\rm R}}} \over g_{*,{\rm f}}} \left({m_\chi/T_{\rm f} \over 15}\right)^4 \left({150 \over m_\chi/T_{\rm R}}\right)^3 \times \left({3 \times 10^{-26} ~ {\rm cm}^3 ~ {\rm s}^{-1} \over \langle \sigma_{\rm ann} v \rangle_{\rm f}}\right) ,
\ee
where $g_{*,{\rm f}}$ is the number of relativistic degrees of freedom at $T_{\rm f}$, 

The contribution from freeze-out must not exceed the observed DM abundance $\Omega_{\chi} h^2 = 0.120 \pm 0.001$~\cite{planckdm}. This, in combination with Eq.~(\ref{treh}) and Eq.~(\ref{tinst}), results in:
\be \label{fobound}  
{H_{\rm f} \over H_{\rm R}} \gtrsim 4 \times 10^{-2}  \left(g_{*,{\rm R}} ~~ g_{*,{\rm f}}\right)^{-1/3} \left({m_\chi \over T_{\rm f}}\right)^{4/3}\times  \left({3 \times 10^{-26} ~ {\rm cm}^3 ~ {\rm s}^{-1} \over \langle \sigma_{\rm ann} v \rangle_{\rm f}}\right)^{4/3} .
\ee
Since $H_{\rm dom} > H_{\rm f}$, and after using $m_\chi \gtrsim 5 T_{\rm f}$, we arrive at the following relation:
\be \label{foemdbound} 
{H_{\rm dom} \over H_{\rm R}} \gtrsim 4 \times 10^{-2} \left(g_{*,{\rm R}} ~ g_{*,{\rm f}}\right)^{-1/3} \times \left({3 \times 10^{-26} ~ {\rm cm}^3 ~ {\rm s}^{-1} \over \langle \sigma_{\rm ann} v \rangle_{\rm f}}\right)^{4/3}.
\ee
\vskip 2mm
\noindent
{$\bullet$ {\bf Freeze-in during EMD:}} 
If $\langle \sigma_{\rm ann} v\rangle_{\rm f}$ is very small, then DM particles will never reach thermal equilibrium at $T > m_\chi$. In this scenario, the DM relic abundance is due to freeze-in of DM production from annihilaitons of the standard model (SM) particles. The main contribution arises from production at $T \sim m_\chi/4$~\cite{GKR}. DM particles produced at higher tempeatures are quickly diluted by the Hubble expansion (when annihilation rate has none or mild dependence on the temperature), while production at lower temperatures is Botzmann suppressed. The DM relic abundance due to freeze-in is given by~\cite{GKR}:
\be \label{fidens}
\Omega_{\chi} h^2 \simeq 0.062 {g^{3/2}_{*,{\rm R}} \over g^3_{*}(m_\chi/4)} \left({150 \over m_\chi/T_{\rm R}}\right)^5 \left({T_{\rm R} \over 5 ~ {\rm GeV}}\right)^2 \times \left({\langle \sigma_{\rm ann} v \rangle_{\rm f} \over 10^{-36} ~ {\rm cm}^3 ~ {\rm s}^{-1}}\right) ,
\ee
For a given DM mass, when the number density of DM particles produced via freeze-in becomes comparable to that from freeze-out, it signals a transition  between the two regimes. Then the annihilation rate at which the transition occurs can be roughly estimated by setting the expressions in Eq.~(\ref{fodens}) and Eq.~(\ref{fidens}) equal. However, for an accurate calculation of this transition one needs to solve a set of Boltzmann equations that also include details of the thermalization of DM particles (including their kinetic equilibrium) and other species with sizeable interactions with DM must be taken into account. The value of $\langle \sigma_{\rm ann} v \rangle_{\rm f}$ at transition depends on $m_\chi$ and $T_{\rm R}$, which is typically within the $10^{-33}-10^{-32}$ cm$^3$ s$^{-1}$ range.

Requiring that freeze-in contribution does not overproduce DM, and after using Eq.~(\ref{treh}) and Eq.~(\ref{tinst}), we find:
\be \label{fibound}
{H(T = m_\chi/4) \over H_{\rm R}} \gtrsim 4 \times 10^3 \left(g_{*,{\rm R}} ~ g^5_*(m_\chi/4)\right)^{-1/7}  \left({m_\chi \over 5 ~ {\rm GeV}}\right)^{8/7} \times \left({\langle \sigma_{\rm ann} v \rangle_{\rm f} \over 10^{-36} ~ {\rm cm}^3 ~ {\rm s}^{-1}}\right)^{4/7} .
\ee
Since $H_{\rm dom} > H(T = m_\chi/4)$, and for $m_\chi > 5$ GeV, this results in the following relation:
\be \label{fiemdbound}
{H_{\rm dom} \over H_{\rm R}} \gtrsim 4 \times 10^3 \left(g_{*,{\rm R}} ~ g^5_*(m_\chi/4)\right)^{-1/7} \times \left({\langle \sigma_{\rm ann} v \rangle_{\rm f} \over 10^{-36} ~ {\rm cm}^3 ~ {\rm s}^{-1}}\right)^{4/7} .
\ee

Some comments are in order before closing this section. First, the fact that satisfying the DM relic abundance from EMD sets a lower bound on its duration can be understood intuitively. For $\langle \sigma_{rm ann} v \rangle_{\rm f} < 3 \times 10^{-26}$ cm$^3$ s$^{-1}$, entropy produced at the end of EMD brings down the relic abundance to its correct value. A longer EMD phase results in a larger entropy release, and hence obtaining enough dilution requires a minimum duration of this epoch.       

Next, we note that Eqs.~(\ref{decbound},\ref{foemdbound},\ref{fiemdbound}) provide absolute lower bounds on $H_{\rm dom}/H_{\rm R}$. The bounds can become stronger when values of $T_{\rm R}$, $m_\chi$, and $\langle \sigma_{\rm ann} v \rangle_{\rm f}$ are specified. Also, while the limits from freeze-out and freeze-in mostly depend on the DM parameters $\langle \sigma_{\rm ann} v \rangle_{\rm f}$ and $m_\chi$, that from decays is mainly dependent on the parameters of the scalar field that drives the EMD phase, namely ${\rm Br}_{\phi \rightarrow \chi}$ and $\alpha_0$, which can broadly vary for differet particle physics realizations of EMD.  

Finally, in the freeze-out scenario Eq.~(\ref{decbound}) and Eq.~(\ref{foemdbound}), and for the freeze-in scenario Eq.~(\ref{decbound}) and Eq.~(\ref{fiemdbound}) must be satisfied simultaneously in order not to overproduce DM during the EMD epoch. In each case, the minimum duration of the EMD phase is set by the larger of the corresponding lower bounds. We note that the stronger limit is in general set by Eq.~(\ref{decbound}). Taking $\langle \sigma_{\rm ann} v \rangle_{\rm f} \sim {\cal O}(10^{-33}-10^{-32})$ cm$^3$ s$^{-1}$, where transition from freeze-in to freeze-out typically happens, maximizes the right-hand side of Eq.~(\ref{foemdbound}) and Eq.~(\ref{fiemdbound}). However, these maximum values are still many orders of magnitude smaller than the right-hand side of Eq.~(\ref{decbound}) unless $\alpha_0$ and/or ${\rm Br}_{\chi \rightarrow \phi}$ are extremely small. This implies that satisfying the lower bound in~(\ref{decbound}) is generally sufficient to meet the DM relic abundance requirement.

\section{Connection to Inflationary Observables}

A remarkable success of the inflationary paradigm is that it provides a natural mechanism for generating the almost scale invariant perturbations. The exact predictions for the values of the scalar spectral index $n_s$ and tensor-to-scalar ratio $r$ depends on the specifics of the model. The predictions for these quantities depends on the model of inflation (usually specified by the inflaton potential) and the number of e-foldings between horizon exit of cosmologically relevant perturbations and the end of inflation. In the presence of an era of EMD, the number of e-foldings of inflation between the time when the pivot scale $k_* = 0.05$ Mpc$^{-1}$ left the horizon and the end of inflation can be written as~ \cite{LL, RG}: 
\bel{nkt}
N_{k_*} \approx 57.3 + {1 \over 4} \ln r - \Delta N_{\rm reh} - \Delta N_{\rm EMD} ,
\ee
where
\be \label{nkt2}
\Delta N_{\rm reh} \equiv {1 - 3 w_{\rm reh} \over 6 (1 + w_{\rm reh})} log \left({H_{\rm inf} \over H_{\rm reh}}\right) ~ ~ ~ , ~ ~ ~ \Delta N_{\rm EMD} \equiv 
 {1 \over 6} \left({H_{\rm dom} \over H_{\rm R}}\right).
\ee
Here $H_{\rm inf}$ is the Hubble rate duting inflation, $H_{\rm reh}$ is the Hubble rate when the universe becomes RD after inflation for the first time, and $H_{\rm dom}$ and $H_{\rm R}$ (discussed in the previous section) denote the Hubble rate at the beginning and the end of EMD epoch respectively (where $H_{\rm dom} < H_{\rm reh}$).   

The first two terms on the right-hand side of Eq.~(\ref{nkt}) define a canonical value for $N_{k_*}$ in a standard thermal history where the universe becomes RD immediately after inflation ends. However, the dynamics of reheating after inflation (for reviews, see~\cite{Review1}) is in general significantly more complex and may involve various non-perurbative and perturbative processes that eventually yield a thermal bath of elementary particles in full equilibrium. The third term on the right-hand side of Eq.~(\ref{nkt}) takes this into account with $w_{\rm reh}$ determining the effective equation of state of the universe during transition from inflation to RD. The last term on the right-hand side of (\ref{nkt}) represents the effect of an epoch of EMD on $N_{k_*}$.     
  
Attributing the entire allowed change in $N_{k^*}$ from its canoncial value to an EMD era results in a conservative lower bound on $H_{\rm dom}/H_{\rm R}$ through Eq.~(\ref{nkt}) and Eq.~(\ref{nkt2}). First, theoretical arguments and numerical simulations suggest that generally $0 \leq w_{\rm re} \leq 1/3$~\cite{PFKP}, which implies that inflationary reheating typically results in $\Delta N_{\rm reh} > 0$. Also, there could be multiple phases of EMD (which, for example, are driven by multiple moduli in explicit string coinstrucions~\cite{ACDS}). While each EMD phase reduces $N_{k_*}$ relative to its canoncial value, only the last one is typically relevant for non-thermal DM productions.

To quantify the connetion between EMD and inflationary observables, we consider the following parametrization of $n_s$ and $r$ in terms of $N_{k_*}$:}
\be \label{param}
n_s \simeq 1 - {a \over N_{k_*}} ~ ~ ~ , ~ ~ ~ r \simeq {b \over N^c_{k_*}} .
\ee
These relations hold for a large number of inflationary models that are consistent with the latest Planck results~\cite{planckinflation}. This includes two important universality classes of single field models discussed in~\cite{Roest:2013fha}.

Class I models satisfy $a = c$ and $b \sim {\cal O}(10)$. The prototypical models include Starobinsky inflation~\cite{Starobinsky:1980te} and Higgs inflation~\cite{Bezrukov:2007ep} for which $a = 2$ and $b \sim 12. $
Class I models have a small scalare-to-tensor ratio $r \lesssim {\cal O}(0.01)$. 

Class II models in this class are characterized by $b = 8 (a -1)$ and $c = 1$. The large field models of inflation fall within this class. The primary examples are the monomial potentials $V(\varphi) \propto \varphi^{2 (a -  1)}$ ($a = 2$ for the quadratic model \cite{Linde} and $ V(\varphi) \sim \varphi, \varphi^{2/3}$ arise in the axion monodromy models \cite{monodromy}). Class II models have a sizable scalar-to-tensor ratio $r \sim {\cal O}(0.1)$.

Also, a new class of inflationary attractor models (called $\alpha$-attractor) has been recently discussed~\cite{Kallosh:2013yoa}. These models smoothly interpolate between the models in the above mentioned universality classes (barring a few models) as the parameter $\alpha$ is varied. For large values of $\alpha$, these models are reduced to the monomial models of Class II. On the other hand, for $\alpha = 1$, the models predict values of $n_s$ and $r$ akin to those in Starobinsky inflation or Higgs inflation of Class I. For smaller values of $\alpha$, these models predict a very small $r$. An important example is K\"ahler moduli inflation \cite{Conlon:2005jm}, which is obtained for $\alpha \sim 3 \times 10^{-8}$. The relations in~(\ref{param}) hold for this model 
with $a = 2$, $c = 3$, and $b \sim 10^{-4}$. A specific model in supergravity with $\alpha = 1/9$ has been proposed by Goncharov and Linde \cite{Goncharov:1983mw}. 

With the help of Eq.~(\ref{param}), we can find the range $N^{\rm min}_{k_*} \leq N_{k^*} \leq N^{\rm max}_{k_*}$ that corresponds to the observationally allowed range $n^{\rm max}_s \leq n_s \leq n^{\rm max}_s$:
\be \label{Nrange}
N^{\rm min}_{k_*} = {a \over 1 - n^{\rm min}_s} ~ ~ ~ , ~ ~ ~ N^{\rm max}_{k_*} = {a \over 1 - n^{\rm max}_s} ~.
\ee
Then for $w_{\rm reh} \leq 1/3$, which is typically the case as mentioned above, Eq.~(\ref{nkt}) implies that:
\be \label{DeltaEMD}
\Delta N_{\rm EMD} \lesssim 57.3 - N^{\rm min}_{k_*} + {1 \over 4} ~ {\rm ln} r(N^{\rm min}_{k_*}). 
\ee
For a specific model of inflation, we can use Eq.~(\ref{Nrange}) to find $N^{\rm min}_{k_*}$ and Eq.~(\ref{param}) to determine the value of $r(N^{\rm min}_{k_*})$. Substituting these in Eq.~(\ref{DeltaEMD}) will then give the upper bound on $\Delta N_{\rm EMD}$. We note that larger values of $N^{\rm min}_{k_*}$ and/or smaller values of $r$ yield a stronger upper limit on $\Delta N_{\rm EMD}$.

\section{Constraints on Non-thermal Dark Matter}

In this section, we combine DM considerations and limits from CMB experiments, to constrain non-thermal DM from an epoch of EMD. According to the latest Planck results~\cite{planckinflation}, the scalar spectral index $2 \sigma$ allowed range for the $\Lambda$CDM+$r$ model from Planck data alone and from Planck plus BK14 and BAO data is given by $n_s = 0.9659 \pm 0.0082$ and $n_s = 0.9670 \pm 0.0074$ respectively. Future CMB experiments are expected to shrink the error bar on $n_s$ by a factor of $\sim 2$~\cite{next1}. 

In Table.~1, we show the values of $N^{\rm min}_{k_*}$, $r(N^{\rm min}_{k_*})$, and the upper bound on $\Delta N_{\rm EMD}$ for representative models of inflation that are compatible with the latest Planck results. We then use Eqs.~(\ref{nkt2}) and Eq.~(\ref{DeltaEMD}) to obtain the following inequaltiy:
\be \label{boundEMD}
{\rm ln}\left({H_{\rm dom} \over H_{\rm R}}\right) \lesssim 344 - 6 N^{\rm min}_{k_*} + {3 \over 2} ~ {\rm ln}r(N^{\rm min}_{k_*}) ,
\ee
and apply Eq.~(\ref{decbound}) to its left-hand side to translate the information in Table.~1 to find constraints in the $\alpha_0-{\rm Br}_{\phi \rightarrow \chi}$ plane. As mentioned before, the lower bounds on $H_{\rm dom}/H_{\rm R}$ from freeze-out of Eq.~(\ref{foemdbound}) and freeze-in of Eq.~(\ref{fiemdbound})) are typically much weaker than that from decays of Eq.~(\ref{decbound}), and thus satisfy Eq.~(\ref{boundEMD}) for the entire relevant range of $\langle \sigma_{\rm ann} v \rangle_{\rm f}$. 

\begin{table}
\label{table}

\begin{tabular}{|p{5.8cm}|p{5.0cm}|p{5.2cm}| }
\hline
 & PLANCK18  & PLANCK18 + BK14 + BAO  \\
\hline
\bf{Inflation Models}
&  
\begin{tabular} {p{1.0cm}|p{1.6cm}|p{2cm} } 
$N_{k_*}^{min}$ & $r (N_{k_*}^{min})$ & $\Delta N_{EMD}^{up} $ \\ 
\end{tabular} 
& 
\begin{tabular} {p{1.0cm}|p{1.6cm}|p{2cm} } 
$N_{k_*}^{min}$ & $r (N_{k_*}^{min})$ & $\Delta N_{EMD}^{up} $\\ 
\end{tabular}  \\
\hline
$V({\phi}) \sim \phi^{4/3}$ 
&  
\begin{tabular} {p{1.0cm}|p{1.6cm}|p{2cm} } 
$39.4$ & $0.13$ & $ 17.4$ \\ 
\end{tabular} 
& 
\begin{tabular} {p{1.0cm}|p{1.6cm}|p{2cm} } 
$41.2$ & $0.13$ & $15.5$ \\ 
\end{tabular}  \\
\hline
$V({\phi}) \sim \phi$ 
&  
\begin{tabular} {p{1.0cm}|p{1.6cm}|p{2cm} } 
$35.5$ & $0.11$ & $ 21.3$ \\ 
\end{tabular} 
& 
\begin{tabular} {p{1.0cm}|p{1.6cm}|p{2cm} } 
$37.1$ & $0.11$ & $19.6$ \\ 
\end{tabular}  \\
\hline
Starobinsky/Higgs \! Inflation 
&  
\begin{tabular} {p{1.0cm}|p{1.6cm}|p{2cm} } 
$47.3$ & $0.0054$ & $ 8.7$ \\
\end{tabular} 
& 
\begin{tabular} {p{1.0cm}|p{1.6cm}|p{2cm} } 
$49.5$ & $0.0049$ & $6.5$ \\ 
\end{tabular}  \\
\hline
%
%
K\"ahler Moduli Inflation 
&  
\begin{tabular} {p{1.0cm}|p{1.6cm}|p{2cm} } 
$47.3$ & $ 9.46\times10^{-10}$ & $ 4.8$ \\ 
\end{tabular} 
& 
\begin{tabular} {p{1.0cm}|p{1.6cm}|p{2cm} } 
$49.5$ & $8.24\times 10^{-10}$ & $ 2.57 $ \\ 
\end{tabular}  \\
\hline
%
%
Goncharov-Linde \! Model ($\alpha = 1/9$)
&  
\begin{tabular} {p{1.0cm}|p{1.6cm}|p{2cm} } 
$47.1$ & $0.00059$ & $ 8.1$ \\ 
\end{tabular} 
& 
\begin{tabular} {p{1.0cm}|p{1.6cm}|p{2cm} } 
$49.3$ & $ 0.00054$& $5.8$ \\ 
\end{tabular}  \\
\hline
\end{tabular}
\begin{flushleft}
\caption{The values of $N^{\rm min}_{k_*}$, $r(N^{\rm min}_{k_*})$, and the upper bound on $\Delta N_{\rm EMD}$ for representative inflation models that satisfy Eq.~(\ref{param}) and are compatible with the latest PLANCK results~\cite{planckinflation}. 
The left and right columns correspond to the $2 \sigma$ range for $n_s$ allowed by PLANCK data alone and by PLANCK plus BK14 and BAO data respectively. 
}
\end{flushleft}
\end{table} 

\begin{figure}[htbp]
  \begin{minipage}[b]{0.48\linewidth}
    \centering
    \includegraphics[width=\linewidth]{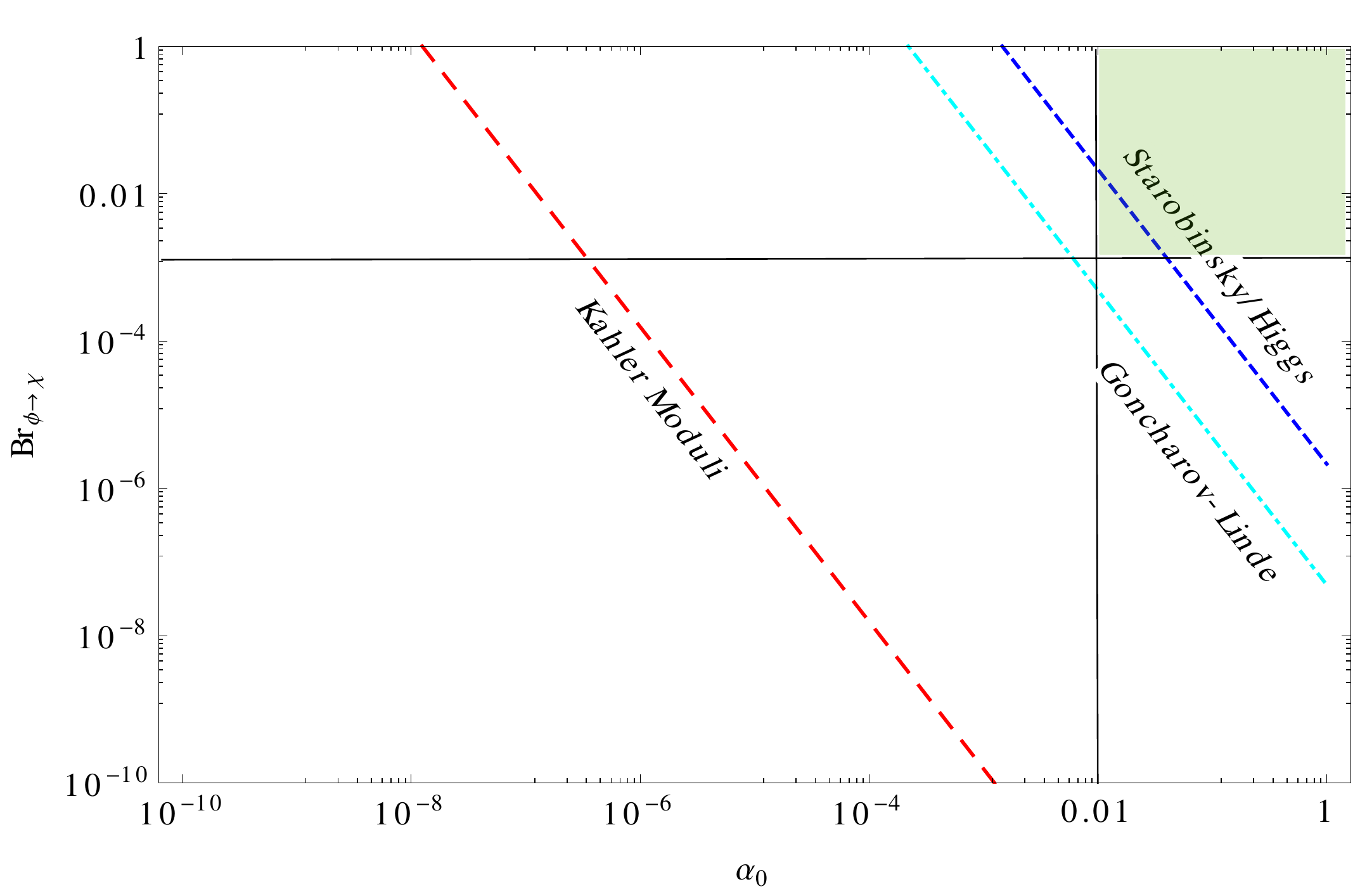}
    \label{fig:chapter001_dist_001}
  \end{minipage}
  \hspace{0.5cm}
  \begin{minipage}[b]{0.48\linewidth}
    \centering
    \includegraphics[width=\linewidth]{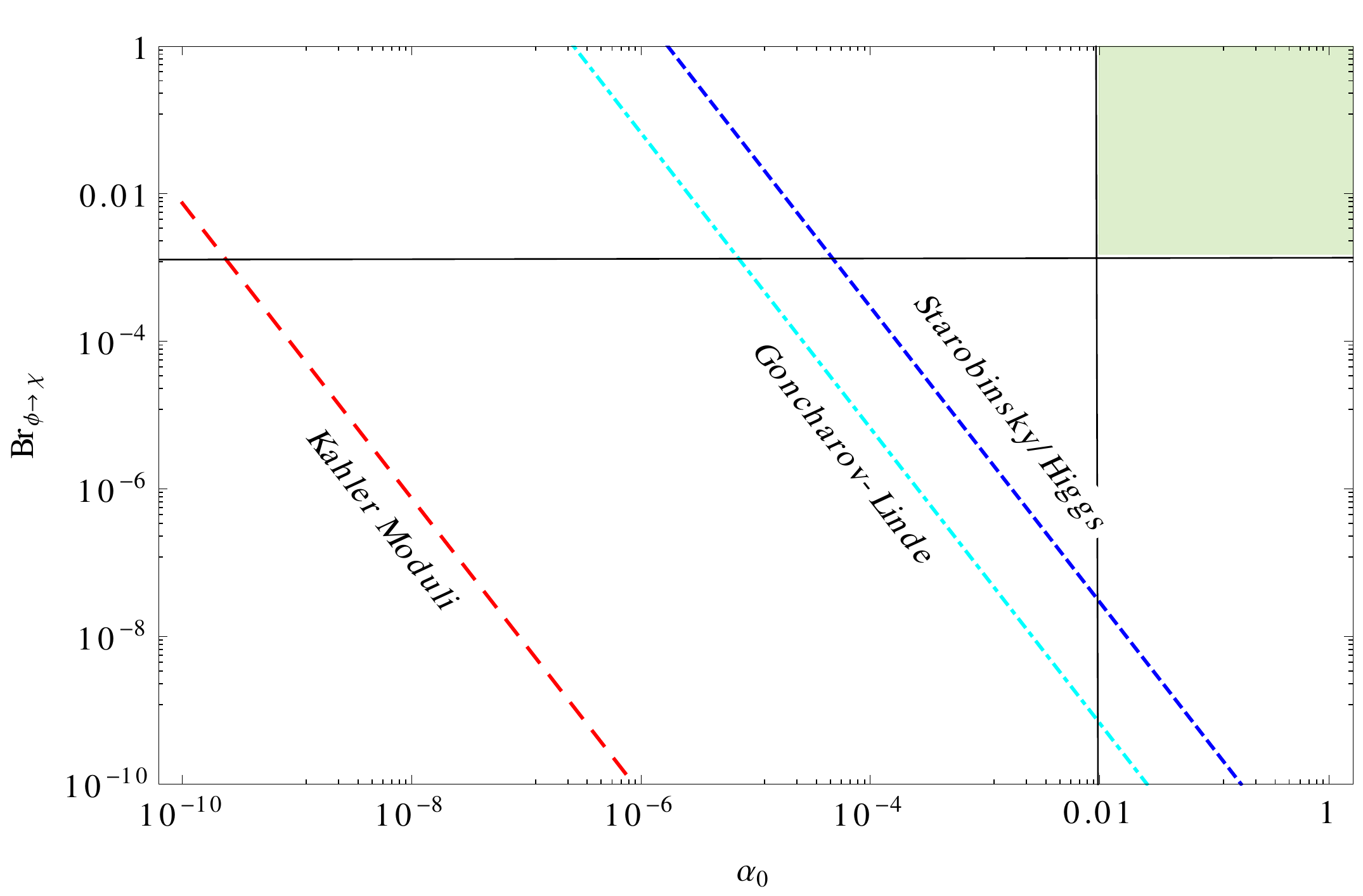}
    \label{fig:chapter001_reward_001}
  \end{minipage}
  \caption{\small{The allowed regions in the $\alpha_0$-${\rm Br}_{\phi \rightarrow \chi}$ plane for inflationary models of Table.~1. The left and right panels correspond to the $2 \sigma$ range for $n_s$ allowed by Planck data alone and PLANCK plus BK14 and BAO data respectively. Each model is represented by a line, and part of the plane to the right of each line is disallowed. The entire plane is allowed for monodromy models.
The shaded region corresponds to the typical parameter space for a modulus-driven EMD phase.}}
\end{figure}

In Fig.~1, we include inflationary models in Table.~1 in the $\alpha_0-{\rm Br}_{\phi \rightarrow \chi}$ plane. Each model is represented by a line and part of the plane to the right of that line is disallowed by the latest Planck results. Data from the future CMB experiments~\cite{next1} will expectedly move these lines to the left and thereby result in tighter constraints. We note the following interesting observations.  

First, Planck results only constrain models with $r \lesssim {\cal O}(0.01)$, while models that have a relatively large $r$ (namely axion monodromy models) do not show up in the figure. This can be understood from Eq.~(\ref{boundEMD}) where a smaller $r$ lowers its right-hand side, and hence makes the inequality stronger. On the other hand, the inequality can be satisfied easily for larger values of $r$ and/or smaller values of $N^{\rm min}_{k_*}$ (as is case for axion monodromy models).

Second, the disallowed parts of the parameter space include the typical region for supersymmetric DM from a modulus-driven EMD era (the shaded area in Fig. 1). Generic arguments based on effective field theory estimates~\cite{mismatch, DRT} or explicit calculations \cite{Cicoli:2016olq} give the amplitude of $\phi$ at the onset of its oscillations to be $\gtrsim {\cal O}(0.1 M_{\rm P})$, implying that $\alpha_0 \gtrsim {\cal O} (10^{-2})$. Also, the branching fraction of moduli decay to supersymmetric DM typically varies within the range ${\cal O}(10^{-3}) \lesssim {\rm Br}_{\phi \rightarrow \chi} \lesssim {\cal O}(1)$, where the exact value depends on the explicit string construction and the lower bound is set by the three-body decay of moduli~\cite{ADS}.

As we see in Fig.~1, parts of the parameter space where $\alpha_0 \ll 1$ and/or ${\rm Br}_{\phi \rightarrow \chi} \ll 10^{-3}$ are allowed in most of the cases. These conditions can be accommodated in situations where a scalar field in the visible sector drives an EMD phase. A notable example is non-thermal DM from late decay of supersymmetric flat direction~\cite{EM}. The initial amplitude of $\phi$ in this case is set by the higher order operators that lift the flatness of its potential, and $\alpha_0 \ll 1$ is in general possible~\cite{DRT}. Situations with ${\rm Br}_{\phi \rightarrow \chi} \ll 10^{-3}$ are also possible when $\phi$ belongs to the visible sector.  For example, this can be achieved in the model in~\cite{ADS2} where $\phi$ is a visible sector singlet. In this model, DM has no direct couplings to $\phi$. As a result, at the leading order, DM production from $\phi$ decay occurs at the one-loop (for Bino-type DM) or two-loop (for Higgsino type DM) level. A combination of loop factors and kinematic suppression can result in ${\rm Br}_{\phi \rightarrow \chi} \sim {\cal O}(10^{-8})$ or smaller.

Finally, as the future CMB experiments expected to push the allowed parameter space to the lower-left corner of the $\alpha_0-{\rm Br}_{\phi \rightarrow \chi}$ plane, we may need to take the contribution from freeze-out/in during EMD to DM production into account. It is seen from Eqs.~(\ref{decbound},\ref{foemdbound},\ref{fiemdbound}) that the lower bounds on $H_{\rm dom}/H_{\rm R}$ from freeze-out/in can become comparable to that from direct decay for $\alpha^2_0 ~ {\rm Br}_{\phi \rightarrow \chi} \lesssim 10^{-25}$. Inclusion of the freeze-out/in contribution to the DM relic abundance will make the constraints that we have obtained here stronger.

\section{Discussions and Conclusion}

We would like to emphasize that the constraints depicted in Fig.~1 are on the conservative side. First, the upper bounds on $H_{\rm dom}/H_{\rm R}$ in Eq.~(\ref{boundEMD}) are obtained assuming that the universe enters a RD phase right after inflation. Also, Eq.~(\ref{decbound}) gives a model-independent absolute lower bound on $H_{\rm dom}/H_{\rm R}$ in order to satisfy the DM relic abundance. Including model details of reheating after inflation or considering specific particle physics realizations of the EMD epoch can make the corresponding inequalities stronger and therby lead to (much) tighter constraints. 

We would also like to note that the CMB limits on $n_s$ used here are for the $\Lambda$CDM+$r$ model. Both the mean value and $2 \sigma$ error of $n_s$ in extensions of this model will be different, which can affect our constraints. Notably, inclusion of dark radiation results in the $2 \sigma$ allowed range $n_s = 0.9607^{+0.0176}_{-0.0168}$ for Planck data alone and $n_s = 0.9660 \pm 0.0140$ for Planck plus BK14 and BAO data~\cite{planckinflation}. The resulting $N^{\rm min}_{k_*}$ from Eq.~(\ref{Nrange}), hence the upper bounds on the duration of EMD from Eq.~(\ref{boundEMD}), will be significantly weaker in this case. Thus future CMB data~\cite{next1} will likely be needed in order to find constraints comparable to those in Fig.~1.        

In conclusion, we studied viability of non-thermal DM from a period of EMD in light of CMB data. Motivated by the increasingly tighter upper limits from indirect searches on the DM annihilation rate, we focused on the case with small annihilation rate $\langle \sigma_{\rm ann} v \rangle_{\rm f} < 3 \times 10^{-26}$ cm$^3$ s$^{-1}$. We found interesting constraints on the parameter space of the EMD phase by combining the lower and upper bounds on its duration from the DM relic abundance consideration and the CMB data on the scalar spectral index respectively. In particular, inflationary models with $r \lesssim {\cal O}(0.01)$ that are compatible with the latest Planck results may disfavor non-thermal supersymmetric DM from a modulus-driven EMD. This conclusion holds as long as the post-inflationary universe has an equation of state characterized by $w \leq 1/3$, and can become stronger with data from the future CMB experiments.

\section*{Acknowledgements}

We are thankful to Michele Cicoli for useful discussions. The work of R.A. is supported in part by NSF Grant No. PHY-1720174. Both K.D. and A.M. are partially supported by Ramanujan Fellowships sponsored by SERB, Department of Science and Technology, Govt. of India. R.A. and K.D. would like to express a special thanks to the Mainz Institute for Theoretical Physics (MITP), where this work started, for its hospitality and support. K.D. and A.M. would like to thank the Abdus Salam International Centre for Theoretical Physics, Trieste for hospitality when the final stages of the work were completed.


\begin{thebibliography}{99}

\bibitem{BHS}
G. Bertone, D. Hooper and J. Silk, Phys. Rept. {\bf 405}, 279 (2005) [e-Print: hep-ph/0404175]. 


\bibitem{fermi1}
M. Ackermann {\it et al.} [Fermi-LAT Collaboration], Phys. Rev. Lett. {\bf 115}, 231301 (2015) [e-Print: arXiv:1503.02641 [astro-ph.HE]]. 


\bibitem{fermi2}
A. Albert {\it et al.} FERMI-LAT and DES Collaborations], Astrophys. J. {\bf 834}, 110 (2017) [e-Print: arXiv:1611.03184 [astro-ph.HE]]. 


\bibitem{Beacom}
R. K. Leane, T. R. Slatyer, J. F. Beacom and K. C. Y. Ng, e-Print: arXiv:1805.10305 [hep-ph]. 


\bibitem{KT}
M. Kamionkowski and M. S. Turner, Phys. Rev. D {\bf 42}, 3310 (1990).



\bibitem{GKR}
G. F. Giudice, E. W. Kolb and A. Riotto, Phys. Rev. D {\bf 64}, 023508 (2001);
A. L. Erickcek, Phys. Rev. D {\bf 92}, 103505 (2015) [e-Print: arXiv:1504.03335 [astro-ph.CO]].



\bibitem{KMY}
M. Kawasaki, T. Moroi and T. Yanagida, Phys. Lett. B {\bf 370}, 52 (1996) [e-Print: hep-ph/9509399]. 



\bibitem{GG}
G. B. Gelmini and P. Gondolo, Phys. Rev. D {\bf 74}, 023510 (2006) [e-Print: hep-ph/0602230].

\bibitem{ADS}
R. Allahverdi, B. Dutta and K. Sinha, Phys. Rev. D {\bf 83}, 083502 (2011) [e-Print: arXiv:1011.1286 [hep-ph]].


\bibitem{KSW}
G. Kane, K. Sinha and S. Watson, Int. J. Mod. Phys. D {\bf 24}, 1530022 (2015).
 
\bibitem{cmp}
G.~D.~Coughlan, W.~Fischler, E.~W.~Kolb, S.~Raby and G.~G.~Ross,
  Phys.\ Lett.\ B {\bf 131}, 59 (1983);
T.~Banks, D.~B.~Kaplan and A.~E.~Nelson,
  Phys.\ Rev.\ D {\bf 49}, 779 (1994) [e-Print: hep-ph/9308292];
B.~de Carlos, J.~A.~Casas, F.~Quevedo and E.~Roulet,
  Phys.\ Lett.\ B {\bf 318}, 447 (1993) [e-Print: hep-ph/9308325].

  
\bibitem{LL}
A. R. Liddle and S. M. Leach, Phys. Rev. D {\bf 68}, 103503 (2003) [e-Print: astro-ph/0305263];
  K.~Dutta and A.~Maharana,
  Phys.\ Rev.\ D {\bf 91}, 043503 (2015)
  [e-Print: arXiv:1409.7037 [hep-ph]].




\bibitem{EGOW} 
  R.~Easther, R.~Galvez, O.~Ozsoy and S.~Watson,
  Phys.\ Rev.\ D {\bf 89}, 023522 (2014)
  [e-Print: arXiv:1307.2453 [hep-ph]].
\bibitem{DDM} 
  K.~Das, K.~Dutta and A.~Maharana,
  Phys.\ Lett.\ B {\bf 751}, 195 (2015)
  [e-Print: arXiv:1506.05745 [hep-ph]].

  
  
\bibitem{related}
  P.~Cabella, A.~Di Marco and G.~Pradisi,
  Phys.\ Rev.\ D {\bf 95}, 123528 (2017)
  [e-Print: arXiv:1704.03209 [astro-ph.CO]];
  S.~Bhattacharya, K.~Dutta and A.~Maharana,
  Phys.\ Rev.\ D {\bf 96}, 083522 (2017) 
   Addendum: [Phys.\ Rev.\ D {\bf 96}, 109901 (2017)]
  [e-Print: arXiv:1707.07924 [hep-ph]];
  A.~Maharana and I.~Zavala,
  Phys.\ Rev.\ D {\bf 97}, 123518 (2018)
  [e-Print: arXiv:1712.07071 [hep-ph]];
  M.~A.~Amin, J.~Fan, K.~D.~Lozanov and M.~Reece,
  arXiv:1802.00444 [hep-ph].
  

\bibitem{infreh}
P. S. Bhupal Dev, A. Mazumdar and S. Qutub, Front. in Phys. {\bf 2}, 26 (2014) [e-Print: arXiv:1311.5297 [hep-ph]];
D. Maity and P. Saha, arXiv:1801.03059 [hep-ph];
D. Maity and P. Saha, arXiv:1804.10115 [hep-ph].   



\bibitem{planckinflation}
Y. Akrami {\it et~al} [PLANCK Collaboration], arXiv:1807.06211 [astro-ph.CO].




\bibitem{next1} 
  K.~N.~Abazajian {\it et al.} [CMB-S4 Collaboration],
  arXiv:1610.02743 [astro-ph.CO].
  F.~Finelli {\it et al.} [CORE Collaboration],
  arXiv:1612.08270 [astro-ph.CO];
  T.~Matsumura {\it et al.},
  J.\ Low.\ Temp.\ Phys.\  {\bf 176}, 733 (2014)
  [e-Print: arXiv:1311.2847 [astro-ph.IM]].



\bibitem{Howie}
H. Baer, K-Y Choi, J. E. Kim and L. Roszkowski, Phys. Rept. {\bf 555}, 1 (2015) [e-Print: arXiv:1407.0017 [hep-ph]]. 



\bibitem{Hooper}
J. A. Dror, E. Kuflik and W. H. Ng, Phys. Rev. Lett. {\bf 117}, 211801 (2016) [e-Print: arXiv:1607.03110 [hep-ph]];
A. Berlin, D. Hooper and G. Krnjaic, Phys. Rev. D {\bf 94}, 095019 (2016) [e-Print: arXiv:1609.02555 [hep-ph]];
J. A. Dror, E. Kuflik, B. Melcher and S. Watson, Phys. Rev. D {\bf 97}, 063524 (2018) [e-Print: arXiv:1711.04773 [hep-ph]].



\bibitem{MR}
T. Moroi and L. Randall, Nucl. Phys. B {\bf 570}, 455 (2000) [e-Print: hep-ph/9906527]; 
B. S. Acharya, G. Kane, S. Watson and P. Kumar, Phys. Rev. D {\bf 80}, 083529 (2009) [e-Print: arXiv:0908.2430 [astro-ph.CO]].



\bibitem{planckdm}
N. Aghanim {\it et al.} [PLANCK Collaboration], e-Print: arXiv:1807.06209 [astro-ph.CO]].

\bibitem{RG}
T. Rehagen and G. B. Gelmini, JCAP {\bf 1506}, 039 (2015) [e-Print: arXiv:1504.03768 [hep-ph]].







 \bibitem{Review1} 
  R.~Allahverdi, R.~Brandenberger, F.~Y.~Cyr-Racine and A.~Mazumdar,
  Ann.\ Rev.\ Nucl.\ Part.\ Sci.\  {\bf 60}, 27 (2010)
  [e-Print: arXiv:1001.2600 [hep-th]];
M. A. Amin, M. P. Hertzberg, D. I. Kaiser and J. Karouby, Int. J. Mod. Phys. D {\bf 24}, 1530003 (2014) [e-Print: arXiv:1410.3808 [hep-ph]]. 



\bibitem{PFKP} 
  D.~I.~Podolsky, G.~N.~Felder, L.~Kofman and M.~Peloso,
  Phys.\ Rev.\ D {\bf 73}, 023501 (2006) [e-Print: hep-ph/0507096];
K. D. Lozanov and M. A. Amin, Phys. Rev. Lett. {\bf 119}, 061301 (2017) [e-Print: arXiv:1608.01213 [astro-ph.CO]].

 


\bibitem{ACDS}
  R.~Allahverdi, M.~Cicoli, B.~Dutta and K.~Sinha,
  Phys.\ Rev.\ D {\bf 88}, 095015 (2013) [e-Print: arXiv:1307.5086 [hep-ph]].


 

\bibitem{Roest:2013fha}
  D.~Roest,
  JCAP {\bf 1401}, 007 (2014)
  [e-Print: arXiv:1309.1285 [hep-th]].



\bibitem{Starobinsky:1980te}
  A.~A.~Starobinsky,
  Phys.\ Lett.\ B {\bf 91}, 99 (1980).

  
\bibitem{Bezrukov:2007ep} 
  F.~L.~Bezrukov and M.~Shaposhnikov,
  Phys.\ Lett.\ B {\bf 659}, 703 (2008)
  [e-Print: arXiv:0710.3755 [hep-th]].
  




%
%
%
%
%
%
%
%
  
\bibitem{Linde}
  A.~D.~Linde,
  Phys.\ Lett.\ B {\bf 129}, 177 (1983);
   V.~A.~Belinsky, I.~M.~Khalatnikov, L.~P.~Grishchuk and Y.~B.~Zeldovich,
  Phys.\ Lett.\ B {\bf 155}, 232 (1985);
  T.\ Piran and R.\ M.\ Williams,
  Phys.\ Lett.\ B {\bf 163}, 331 (1985).


\bibitem{monodromy} 
  L.~McAllister, E.~Silverstein and A.~Westphal,
  Phys.\ Rev.\ D {\bf 82}, 046003 (2010);
  E.~Silverstein and A.~Westphal,
  Phys.\ Rev.\ D {\bf 78}, 106003 (2008).

  
  \bibitem{Kallosh:2013yoa} 
  R.~Kallosh, A.~Linde and D.~Roest,
  JHEP {\bf 1311}, 198 (2013)
  [e-Print: arXiv:1311.0472 [hep-th]].
  

\bibitem{Conlon:2005jm} 
  J.~P.~Conlon and F.~Quevedo,
  JHEP {\bf 0601}, 146 (2006)
  [e-Print: hep-th/0509012].
  
\bibitem{Goncharov:1983mw} 
  A.~B.~Goncharov and A.~D.~Linde,
  Phys.\ Lett.\  {\bf 139B}, 27 (1984).



\bibitem{mismatch}
  M.~Dine, W.~Fischler and D.~Nemeschansky,
  Phys.\ Lett.\ B {\bf 136}, 169 (1984);
  G.~D.~Coughlan, R.~Holman, P.~Ramond and G.~G.~Ross,
  Phys.\ Lett.\ B {\bf 140}, 44 (1984);
  A.~S.~Goncharov, A.~D.~Linde and M.~I.~Vysotsky,
  Phys.\ Lett.\ B {\bf 147}, 279 (1984).



\bibitem{DRT}
M. Dine, L. Randall and S. D. Thomas, Phys. Rev. Lett. {\bf 75}, 398 (1995) [e-Print: hep-ph/9503303]; M. Dine, L. Randall and S. D. Thomas, Nucl. Phys. B {\bf 458}, 291 (1996) 291-326 [e-Print: hep-ph/9507453].  


\bibitem{Cicoli:2016olq} 
  M.~Cicoli, K.~Dutta, A.~Maharana and F.~Quevedo,
  JCAP {\bf 1608}, 006 (2016). [e-Print: arXiv:1604.08512 [hep-th]]
  
\bibitem{EM}
K. Enqvist and J. McDonald, Nucl. Phys. B {\bf 538}, 321 (1999) [e-Print: hep-ph/9803380]. 

\bibitem{ADS2}
R. Allahverdi, B. Dutta and K. Sinha, Phys. Rev. D {\bf 87}, 075024 (2013) [e-Print: arXiv:1212.6948 [hep-ph]].

\end{thebibliography}
\end{document}